\documentstyle[twoside, epsfig]{article}
\input pz.sty

\begin{document}
\PZhead{4}{30}{2010}{29 September}{5 October}

\PZtitle{The progenitor and remnant of the helium nova V445
Puppis}

\PZauth{V.~P. Goranskij$^{1}$, S.~Yu. Shugarov$^{1,2}$, A.~V. Zharova$^{1}$,
P. Kroll$^{3}$, and E.~A. Barsukova$^{4}$}

\PZinst{Sternberg Astronomical Institute, Moscow University,
Universitetsky Ave. 13, Moscow, 199992 Russia}

\PZinst{Astronomical Institute of the Slovak Academy of Sciences,
Tatranska Lomnica, 05960 Slovakia}

\PZinst{Sternwarte Sonneberg, Sternwartestrasse 32, Sonneberg,
D-96515 Germany}

\PZinst{Special Astrophysical Observatory, Russian Academy of
Sciences, Nizhny Arkhyz, Karachai-Cherkesia, 369167 Russia}

\SIMBADobj{V445 Pup}

\PZabstract{ V445 Pup was a peculiar nova with no hydrogen
spectral lines during the outburst. The spectrum contained strong
emission lines of carbon, oxygen, calcium, sodium, and iron. We
have performed digital processing of photographic images of the
V445 Pup progenitor using astronomical plate archives. The
brightness of the progenitor in the $B$ band was 14\fmm3. It was a
periodic variable star, its most probable period being
$0\fday650654 \pm0\fday000011$. The light curve shape suggests
that the progenitor was a common-envelope binary with a spot on
the surface and variable surface brightness. The spectral energy
distribution of the progenitor between 0.44 and 2.2~$\mu$m was
similar to that of an A0V type star.

After the explosion in 2001, the dust was formed in the ejecta,
and the star became a strong infrared source. This resulted in the
star's fading below $20\mm$ in the $V$ band. Our CCD $BVR$
observations acquired between 2003 and 2009 suggest that the dust
absorption minimum finished in 2004, and the remnant reappeared at
the level of 18\fmm5 $V$. The dust dispersed but a star-like
object was absent in frames taken in the $K$ band with the VLT
adaptive optics. Only expanding ejecta of the explosion were seen
in these frames till March 2007. No reddened A0V type star
reappeared in the spectral energy distribution. The explosion of
V445 Pup in 2000 was a helium flash on the surface of a CO-type
white dwarf. Taking into account the results of modern dynamic
calculations, we discuss the possibility of a white-dwarf core
detonation triggered by the helium flash and the observational
evidence for it. Additionally, the common envelope of the system
was lost in the explosion. Destruction in the system and mass loss
from its components exclude the future SN Ia scenario for
V445~Pup. }

\section{INTRODUCTION}

The outburst of V445 Pup was discovered on 30 December 2000 by
Kanatsu (Kato \& Kanatsu 2000). The earliest observation of V445
Pup in the outburst, dated 19 November 2000, was found in the ASAS
archive. At that time, the brightness of the star was 8\fmm8. The
brightness maximum of 8\fmm46 in the $V$ band was reached on 29
November 2000. The first spectroscopic observations in the
outburst by Wagner et al. (2001) showed that the Balmer emission
and He I lines typical of classical novae were not present in the
spectrum of V445 Pup. The spectra were dominated by emission lines
arising from FeII, CaII, CII, NaI, OI. Line widths corresponded to
an expansion velocity of about 1000 km~s$^{-1}$. The ejecta
produced during the outburst allow us to consider V445 Pup as a
nova.

The nature of classical novae is known to be a thermonuclear
explosion of hydrogen on the surface of a white dwarf in a
semidetached binary system. Hydrogen accumulates on the surface of
the white dwarf due to accretion from a donor, usually a red
dwarf. As a result of hydrogen explosion, strong Balmer lines are
observed in the spectra of classical novae. Ashok \& Banerjee
(2003) suggested that V445 Pup was the unique helium nova
predicted theoretically by Kato et al. (1989) and by Iben \&
Tutukov (1994) who considered the case of a degenerate white dwarf
accreting helium from a helium-rich donor. Note that a subclass of
classical novae called helium-nitrogen (He/N) novae was introduced
by Williams (1992). These novae have spectra with strong Balmer
lines, and they also have HeI and HeII lines. CNO elements seen in
the spectra were mixed by accumulated hydrogen envelope from the
surface of the white dwarf through a dredge-up mechanism (for
instance, cf. Glasner \& Livne 2002). The case of helium nova
suggests that the donor is a nucleus of an evolved star that
previously lost its hydrogen envelope due to accretion.

In the outburst of V445 Pup, the decay of brightness by 1\fmm8
continued gradually for 164 days and was followed by a small
rebrightening between 12 May and 21 June 2001. The last
observation of V445 Pup in the outburst was on 11 July 2001 at
visual magnitude 11.5. Then the star faded rapidly and was not
seen in August 2001. On 4 October 2001, no object brighter than
$V$ = 20\mm\ or $I$ = 19\fmm5 was found by Henden et al. (2001) at
the position of V445 Pup. They remark: ``The object is evidently
shrouded in a thick and dense carbon dust shell, in view of the
apparent over-abundance of carbon in ejecta previously observed in
infrared and optical spectra''. Lynch et al. (2001) detected the
infrared radiation in the 3--14~$\mu$m range just 1~month after
the object had been discovered. The spectrum revealed smooth and
featureless continuum, which they treated as a thermal emission of
dust with the temperatures ranged between 280 and 1300~K. They
suggest that this dust was a product of previous outbursts, at
least in part.

\PZfig{8.5cm}{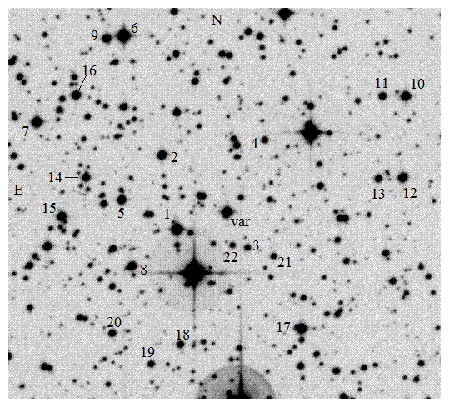}{The finding chart of V445 Pup
and comparison stars. This is a copy of a digitized image obtained
with the UKST Schmidt telescope on 4 April 1980 on IIIaJ emulsion
with a GG 395 filter. The progenitor is indicated as ``var''. $V$
magnitudes, colour indices of marked stars, and corresponding
uncertainties (in units of thousandths of a magnitude) are given
in Table 1.}

The detailed spectral evolution of V445 Pup in the outburst was
studied by Iijima \& Nakanishi (2008). They acquired high- and
medium-resolution spectra for the optical wavelengths 3900--7000
\AA. They confirmed the absence of hydrogen lines and noted
unusually strong emissions of carbon ions. Some metal lines had
P~Cyg profiles with absorption components blue-shifted roughly by
$-500$ km~s$^{-1}$; this velocity was assumed to be the outflow
velocity. The cited authors measured large radial velocity of V445
Pup, +224 $\pm$8 km s$^{-1}$, which suggested that the object
belonged to the old disk population. The distance was estimated
from the interstellar NaI D$_1$ and D$_2$ absorption lines to be
$3.5 < d < 6.5$ kpc; the reddening is $E(B-V) = 0.51$.

\PZfig{10cm}{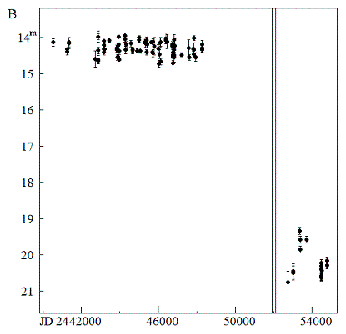}{The pre- and post-outburst light
curve of V445 Pup in the $B$ band. The start and the end of the
2000--2001 outburst are indicated with a double vertical line.}

Lynch et al. (2004) reported that in January, 2004 the object had
faded to fainter than $J = 18$, so that they could not take its
spectrum in the visible range. In the infrared, they detected only
two HeI lines at 1.0830 and 2.0581 $\mu$m, both showing doubled
profiles due to bipolar outflow. The very red continuum was
detected only at $\lambda \ge 1.5$~$\mu$m. It was produced by
emission of hot dust.

Woudt et al. (2009) published the results of post-outburst $JHK$
photometry, adaptive optics imaging in the $K$ band, and
optical-range spectroscopy of V445 Pup. They discovered an
expanding and narrowly confined bipolar shell, the outflow
characterized by large velocity of $6720 \pm650$~km~s$^{-1}$. Some
knots were moving with larger velocities, $8450
\pm570$~km~s$^{-1}$. They derived an expansion parallax distance
of $8.2 \pm0.5$~kpc. They noted that the expansion velocity
measured by Iijima \& Nakanishi (2008) from their high-resolution
spectra in the outburst was only 500~km~s$^{-1}$. Such a big
difference may be due to strong collimation of bipolar ejection
located just in the plane of the sky and inclined to this plane by
only $5\fdeg8{-}3\fdeg7$ (Woudt et al. 2009). The authors assume
that the small inclination angle may confirm the presence of an
orthogonal dust structure closely aligned to the line of sight and
causing the strong extinction observed after the outburst. In
their spatially resolved optical spectrum obtained with the VLT in
January, 2006 in the 4465--7634~\AA\ range, only the emission
lines of [OI], [OII], [OIII], and HeI were seen, but not the
continuum.

The presence of a bright progenitor of V445 Pup having a visual
magnitude of 13.1 was first noted by Platais et al. (2001). Its
absolute proper motion was small, $\mu = 0\farcs008
\pm0\farcs004$. With the distance derived by Woudt et al. (2009),
the luminosity of the progenitor proved to be very large, $\log
L/L_{\odot} = 4.34 \pm0.36$, which is consistent with the absolute
bolometric magnitude $M_{bol} = -6.1 \pm0.9$. Woudt et al. (2009)
note that the derived luminosity suggests that V445 Pup probably
contains a massive white dwarf accreting at a high rate from a
helium star companion. But they did not exclude that the companion
was also a bright star. Liller (2001) reminded of three
hydrogen-deficient cataclysmic variables, CR Boo, CP Eri, and V803
Cen, all of them hot blue objects showing no hydrogen but
revealing HeI emission lines. The absolute magnitude of $-6.1$ is
unprecedented high for a cataclysmic variable, making us to think
about the nature of the progenitor.

The observations of the light curve in the outburst and light
curve modeling by Kato et al. (2008) reveal that the CO white
dwarf in V445~Pup is very massive and close to the Chandrasekhar
mass limit ($M_{\rm wd} \ge 1.35 M_{\odot}$); a half of the
accreted matter remained on the white dwarf after the outburst.
Therefore, V445~Pup was considered a strong candidate for a type
Ia supernova progenitor. P.~Woudt and D.~Steeghs called V445 Pup a
``ticking stellar time bomb'' in the ESO Science Press Release
0943. Taking into account the observations with adaptive optics by
Woudt et al. (2009) which show only spatially resolved products of
eruption but no stellar component, it is hard to maintain the
concept that the mass of the system increases. The question is
what mass of the components was left after the explosion.

The scenario for V445 Pup may be quite different. Recent dynamic
3D simulations by Guillochon et al. (2010) discovered a new
mechanism for the detonation of a core of a sub-Chandrasekhar CO
white dwarf (with a mass lower than $1.4 M_{\odot}$) in a system
with a pure He white dwarf or a He/CO hybrid secondary. Fink et
al. (2010) found that secondary core detonations were triggered
for all of the simulated models ranging in core mass from 0.810 up
to $1.385 M_{\odot}$ with corresponding helium shell masses from
0.126 down to $0.0035 M_{\odot}$. In that paper, the double
detonation scenario remains a potential explanation for type Ia
supernovae. But the destruction of the CO white dwarf means that
V445~Pup, after its outburst in 2000, will not be a type Ia
supernova progenitor. It is of great interest if the narrowly
confined bipolar cones observed by Woudt et al. (2009) are debris
of the detonated white dwarf. The overabundant carbon in the
outburst will also be an evidence for CO white dwarf detonation.

Fortunately, there are many photographic images of V445~Pup in the
world astronomical plate collections suitable for resolving the
puzzle of the progenitor. Woudt et al. (2009) verified the plate
archives at the Harvard--Smithsonian Center for Astrophysics (USA)
and found no prior outbursts in 1897--1955. The progenitor of
V445~Pup was identified on many plates at approximately constant
brightness (from visual comparison with surrounding stars). We
found many plates in archives of the Sternberg Astronomical
Institute (SAI) of the Moscow University (Russia) and in archives
of the Sonneberg Observatory (Germany). Two of us (V.P.G. and
S.Yu.Sh.) performed eye estimates of V445~Pup, V.P.G. for Moscow
plates, and S.Yu.Sh. for Sonneberg plates. The two sets showed
similar behavior and marginal variability. But unexpectedly, the
preliminary frequency analysis revealed the same periodicity in
both sets with the period of 0\fday650653, coinciding to the 6th
significant digit. Both light curves were of low quality.
Therefore we decided to digitize the images of V445~Pup and to
perform digital processing.

\section{OBSERVATIONS AND DATA PROCESSING}

In the Moscow SAI plate collection, we found 51 plates with images
of V445~Pup taken with the SAI Crimean Station 40-cm f/4
astrograph and dated between 15 November 1969 and 4 November 1989.
AGFA ASTRO and ORWO ZU-2 photographic plates, produced in the
former GDR and having high sensitivity in blue light, were used,
the exposure times were 45 minutes. The geographic position of the
SAI Crimean Station is $2\hr16\mm08\sec, +44\deg43\arcm42\arcs$.
The declination of V445~Pup is about $-26^\circ$. This means that
the highest altitude of the star above the horizon is 19\deg. The
observations were limited to a 3-hour visibility time around this
point. The photographic plates were mostly centered at $\tau$~CMa,
they cover an area of 10\deg$\times$10\deg. The region of about
20\arcm $\times$ 20\arcm\ centered at V445 Pup was digitized for
each plate using the SAI CREO EverSmart Supreme scanner. CREO
scanner frames are in the TIFF format.

We found 56 measurable images of V445 Pup on the plates of the
Sonneberg Observatory collection dated between 19 March 1984 and
17 January 1991. These plates were taken with the 40-cm f/4
astrograph having optics basically similar to that of the SAI
Crimean Station astrograph. Also, plates of basically the same
type produced in the former GDR were used, so all our photographic
material is very uniform. The Sonneberg Observatory has the
geographic position $0\hr44\mm46\sec, +50\deg22\arcm39\arcs$, it
is about 5\deg\ to the north by latitude compared to the SAI
Crimean Station. Thus, the star rises only to 14\deg\ for this
geographic point, and its visibility time is less than that for
the Crimean Station. Sky images are evidently affected with
variable atmospheric extinction across the plate field. In these
plates, the star is located near the center. The typical exposure
time for these plates is 20 minutes. Images of V445 Pup were
digitized using the Fuji FinePix F10 CCD camera and an ordinary
biconvex lens. Frames made with this camera are in the JPEG
format. After several experiments, we found that this method of
digitizing gave the quality of measurements near the field center
as good as that of a scanner. To increase the S/N ratio, we
co-added several subsequent frames in night series.
The frames were combined with matching two stars and
with the field rotation. The combined exposures of co-added
frames were between 40 and 60 minutes. The total number of
measurements for the star, including co-added ones, is 31.

Additionally, we measured all the Internet-accessible Digital Sky
Survey images of V445 Pup in $B$, $R$, and $I$ bands and used
2MASS $JHK$ magnitudes to study the spectral energy distribution
of the progenitor.

All the frames were processed in the Windows BITMAP format.
Extraction of images was made using the aperture method with
star-profile correction; the WinPG software developed by V.P.G.
was utilized. Special software was written by V.P.G. to
approximate the characteristic curves with an $n$th-degree
polynomial, with graphical output. Practically, approximations
with $n = 1$ or 2 were optimal. The total number of comparison
stars used to build a characteristic curve was 23; a few stars
with the largest deviations were eliminated from calculations, and
the characteristic curves were re-calculated in such cases. The
r.m.s. deviation of comparison stars from the polynomial fit was
formally taken for the uncertainty of V445 Pup measurement.

Carrying out our photographic measurements, we used the CCD
$BVR_cI_c$ standard sequence in the vicinity of V445 Pup published
by A.~Henden for VSNET. We present the finding chart of the
progenitor in Fig.~1. The standard stars chosen by us are also
marked in this Figure. We give Henden's magnitudes and their
uncertainties for the chosen comparison stars in Table 1 because
they are no longer accessible at the VSNET address.

The Moscow archive observations of the V445~Pup progenitor are
presented in Table~2; the Sonneberg ones, in Table~3; and those
from digitized sky surveys are collected in Table~4.

We performed our observations of the V445 Pup remnant  between 31
March 2003 and 20 October 2009. These observations were acquired
in the Special Astrophysical Observatory (SAO), with the 1-m Zeiss
reflector and CCD $UBVR_CI_C$ photometer equipped with an EEV
42-40 CCD chip. The geographic position of the SAO is
$2\hr45\mm46\sec, +43\deg39\arcm12\arcs$. The highest altitude of
the star over the horizon is 20\deg. This object is difficult for
observations and needs good sky transparency and seeing. Some
constructions of the 6-m telescope dome located to the south of
the 1-m reflector hamper observations of objects with such a
southern declination. Additionally, a part of our observations
were obtained with the SAI Crimean Station's 60-cm reflector and
$UBVR_JI_J$ photometer with the Princeton Instruments VersArray
CCD. Both devices are cooled with liquid nitrogen to a temperature
stabilized at $-130\deg$~C, allowing to record signals from very
faint astrophysical objects. The frames were reduced in the FITS
format. The extraction of images was made using the same aperture
method with star-profile correction, the WinFITS software by
V.P.G. was utilized. Our CCD observations are presented in
Table~5.

\PZfig{15cm}{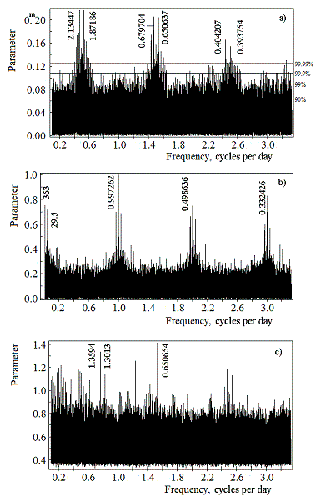}{Periodograms of the V445 Pup
progenitor. (a): The Deeming amplitude spectrum in the
(10--0.3)-day period range. The parameter is the half-amplitude.
(b): The spectral window of the same time series in the
(1000--0.3)-day range. The parameter is the half-amplitude. (c):
The Lafler--Kinman periodogram. The parameter is $\theta^{-1}$,
$\theta$ being the normalized sum of squared magnitude differences
between each two subsequent points of the phased light curve
calculated with a trial period. }

\section{ANALYSIS}

The light curve of V445 Pup in the $B$ band excluding the outburst
is shown in Fig.~2. The variability of the progenitor is evident,
and its full amplitude exceeds the mean uncertainty of the
observations more than thrice.

The frequency analysis of the progenitor observations was
performed using two independent methods: (1) the discrete Fourier
transform for arbitrarily distributed time series (Deeming 1975),
and (2) the phase dispersion minimization (PDM) method (Lafler \&
Kinman 1965). Implementations of these methods are provided with
the EFFECT code developed by V.P.G. We analyzed the combined time
series including Moscow and Sonneberg photographic observations.
The periodograms are shown in Fig.~3a-c. The panels (a) and (b) of
this figure present the amplitude spectrum and the spectral window
for this series. We estimate significance levels for peaks of the
amplitude spectrum using the empirical method suggested by
Terebizh (1992). This method is based on a statistical analysis of
simulated chaotic series generated by mixing the original series.
In the chaotic series, each Julian date gets an accidental
magnitude chosen from the same original series and, as a result,
the chaotic series contains the same magnitudes and times. When we
compute the amplitude spectrum of the chaotic series, we make more
than a million of accidental light-curve implementations with
arbitrary periods and estimate their amplitudes. The software
provides the analysis of the cumulative probability distribution
function for amplitudes in the spectrum of the chaotic series. The
amplitude levels corresponding to cumulative probabilities of 90,
99, 99.9 and 99.99 percent for the chaotic series are plotted in
Fig.~3a as straight lines.

The presence of strong peaks in the amplitude spectrum of the
original series exceeding 99.99 percent amplitude level of the
chaotic series means that the probability of casual appearance of
these peaks is less than 0.01 percent. The progenitor of V445~Pup
was evidently a periodic variable star. The multiplicity of peaks
means that we have multiple solutions for periodicity with the
Moscow and Sonneberg series.

The spectral window (Fig.~3b) demonstrates the periodicity in time
discontinuities in our series amounting to the sidereal day
($P_{sd}=0\fday997262$) and to $P_{sd}/n$, where $n = 2, 3, 4...$.
The amplitudes of these peaks decrease when the period decreases
because of increasing phase window. The phase window for $P_{sd}$
is 0.2. Thus, for each peak in the window spectrum, we have a pair
of symmetrically located alias peaks in the amplitude spectrum.
The light curves corresponding to this pair of peaks have reverse
phase count, so they look as mirror-reflection ones. The list of
periods and frequencies of aliases is given in Table~6. One can
see that the dominating peaks are sidereal-day-related. The
formula of corresponding interdependence is given for each peak in
the last column, `Remark', of Table 6. For $f_0$, we chose the
lowest-frequency wave with the highest amplitude. Peaks of a lunar
month (29.5 day) and of about a year (363 days) in the spectrum of
the window are also present, which are responsible for combs of
small peaks located around the sidereal-day-related peaks in the
star spectrum.

The light curves corresponding to all alias periods are given in
Fig.~4. These are single-wave curves.  The light curves are
approximately sinusoidal, the full amplitude of the sine wave is
about 0\fmm4. Formally, the Deeming method reveals the
highest-amplitude light curves for two periods,
$1\fday871862\pm0\fday00009$ and $2\fday134469 \pm0\fday00011$,
with equal half amplitudes of 0\fmm22. The scatter of all the
light curves reveals essential intrinsic variability. A few points
do evidently contradict the sinusoidal solution. We verified these
points and confirmed their Julian dates and magnitudes. These
light curves may represent the case of reflection effect on the
surface of a secondary star due to heating of a part of its
surface by the X-ray or short-wavelength radiation coming from the
primary star. However, no X-ray source was associated with
V445~Pup before its outburst. In principle, such light curves may
arise due to a large hot spot on the surface of a star.
FK~Com-type stars may be examples of a rotating star having a hot
spot on the surface. These stars are considered to be close binary
systems with a common envelope.

\PZfig{17cm}{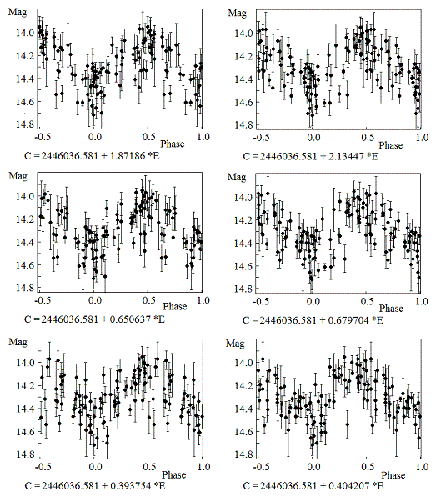}{The light curves plotted for the
periods determined with the Deeming (1975) method and presented in
Table~6. The elements used to calculate phases are given below
each curve.}

Double-wave light curves for periods found by the Deeming method
presented in Table~6 were also calculated and are shown in Fig.~5.
Double-wave light curves are exhibited by W~UMa-type
binaries, these are also stars with common envelopes. In most
observational cases of photometry without additional spectroscopic
information, like radial velocity curves or double lines in the
spectra, we can distinguish between single-wave and double-wave
orbital periods only by differing alternate minima.

\newpage

\PZfig{17cm}{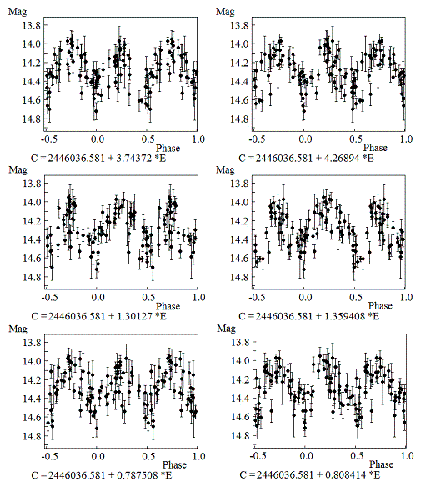}{Light curves plotted for double
periods determined with the Deeming (1975) method and given in
Table~6. The elements used to calculate phases are given below
each curve.}

The difference of minima depths appears due to difference of
surface brightness of the components. Our observations do not show
alternate minima of different depth. Unfortunately, the accuracy
of photographic observations is insufficient to make a reliable
choice between single- and double-wave curves. However, taking
into account that this system contains a massive accreting CO
white dwarf (Kato et al. 2008), we think that a W UMa light curve
is not an acceptable solution because components of such a system
should have different brightness.

The Lafler--Kinman (L--K) method reveals periods not exceeding one
day as preferable: $0\fday650654 \pm0\fday000011$, $0\fday679686
\pm0\fday000012$ day, and their double-wave aliases $1\fday301269
\pm0\fday000044$ and $1\fday359423 \pm0\fday000048$. The
0.650654-day period determined with the L--K method differs
essentially from that determined by the Deeming method because the
light curve plotted with the PDM solution shows a local detail for
the phases between 0.9 and 0.1 that looks like a shallow eclipse
(Fig.~6). However, additional photographic material is needed to
verify if this detail is real. Certainly, a detail like a shallow
eclipse cannot be revealed in a light curve with the Deeming
method. The double-wave light curves found by the L--K method seem
more irregular and asymmetric. Additionally, the light curve with
the period of $0\fday650654 \pm0\fday000011$ is the most symmetric
one and has the lowest dispersion, so we choose it as the best
solution for the present time.

Our investigation shows that one can find a single final solution
for the orbital period of V445 Pup only using observations
acquired at different geographic longitudes, thus increasing the
observational phase window of the sidereal-day period, $P_{sd}$.
Fortunately, there are enough plates in the world plate
collections to get such a solution. It is known that Harvard
plates were taken at the station located in the southern
hemisphere, and this series would have the widest window. We hope
that digitally reduced Harvard plates and Japanese pre-outburst
photographic observations, along with our data, will provide a
true final solution.

\PZfig{8.5cm}{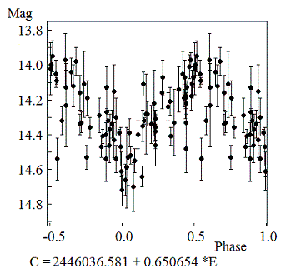}{The light curve for the best
single-wave period found with the Lafler \& Kinman (1965) method.}

\PZfig{8.5cm}{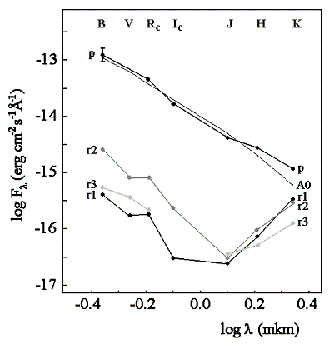}{The spectral energy
distributions determined from photometry. The upper solid curve,
marked $p$, is the energy distribution of the V445 Pup progenitor,
the bars of the $B$ point correspond to the variability amplitude.
The $JHK$ data are from the 2MASS survey. Interstellar extinction
corresponding to $E(B - V) = 0.51$ was taken into account. The
gray curve marked A0 is the energy distribution for an A0V star,
plotted for comparison. The three lower curves are energy
distributions of the V445 Pup remnant measured at different time:
(r1) December 2003; (r2) January 2005; (r3) November 2008. The
$JHK$ observations are from Woudt et al. (2009).}

\section{SPECTRAL ENERGY DISTRIBUTIONS}

We have remeasured $B, R$ and $I$ frames of Digital Sky Surveys
for the V445 Pup progenitor using Henden's photometric
comparison-star sequence. The infrared $JHK$ observations of the
progenitor were taken from the 2MASS survey. Additionally, we
compared these observations to our optical $BVR_CI_C$ observations
of the remnant based on the same comparison-star sequence. The
brightness of the remnant was found to be variable in the
magnitude ranges of $19.3{-}20.8$ in the $B$ band; $18.5{-}20.3$
in the $V$ band; $17.5{-}19.4$ in the $R_C$ band; and $17.7{-}
20.0$ in the $I_C$ band. In Fig.~7, we show the spectral energy
distributions (SEDs) of the progenitor (p) and remnant for three
different dates: December, 2003, in minimum light after the
outburst when its brightness was $V \approx 20.1$ (r1); January,
2005, in the peak of rebrightening at $V \approx 18.6$ (r2); and
November, 2008, after the rebrightening at $V \approx 19.2$ (r3).
To plot the distributions for the remnant, we used  $JHK$
observations by Woudt et al. (2009) from their Table~2, the
closest to our dates. All distributions were dereddened; those for
V445~Pup were dereddened with the colour excess $E(B-V) = 0.51$
(Iijima \& Nakanishi 2008), based on calibration of the
interstellar NaI doublet equivalent width measured in
high-resolution spectra.

We compared our pre-outburst SED with that previously published by
Ashok \& Baner\-gee (2003) who find consistency of the SED with an
accretion disk model. In the $JHK$ bands, they used the same 2MASS
data, but in the optical $B$ and $R$ bands, they used the USNO
A2.0 catalog. The USNO A2.0 $R$ magnitude coincides well with our
photometry based on Henden's comparison stars. But the USNO $B$
value is a half-magnitude brighter than the mean magnitude from
our archive observations and is outside the variability range. In
Fig.~7 by Ashok and Banergee (2003), the $B$ point is located
above the straight line fitting the other points. It is clear that
it is this deviation in the $B$ band that caused the choice of the
accretion disk model. Additionally, Ashok \& Banergee (2003) used
a smaller $E(B{-}V)$ value of 0\fmm25, derived from the extinction
maps by Neckel et al. (1980) and Wooden's (1970) $UBV$ photometry
of stars in the $5\deg\times5\deg$ field around V445 Pup. Woudt et
al. (2009) reanalyzed the problem of interstellar extinction based
on published photometry of open clusters and ascertain that
$E(B{-}V) = 0.51$ can be the lower limit of Galactic reddening,
and the actual value can be between 0.51 and 0.68. They also
include circumstellar reddening in their estimates of the star's
absolute magnitude and luminosity.

Our SED of the V445 Pup progenitor fits that for an A0V star well (A0
in Fig.~7).  If we exclude a possible small excess visible in
the $H$ and $K$ infrared bands, the spectral type may be somewhat
earlier. Probably, this excess is the radiation of dust detected
by Lynch et al. (2001) in the outburst. But this was not the dust
of the surrounding thick disk,  assumed to explain the obscuration
of the whole system after the outburst. No essential infrared
excess is seen in the SED of the progenitor to cause additional
circumstellar light absorption and reddening. It is notable that
the energy distribution of the progenitor is that of a single
star: we do not see any contribution of a cool donor or hot
radiation from an accretion disk. Moreover, Kato et al. (2008)
demonstrated that the progenitor was too bright to be an accretion
disk at the known distance ($3 \le d \le 6.5$ kpc) and with known
Galactic interstellar absorption ($A_V = 1.6$) even for the
largest reasonable mass accretion rate ($10^{-6} M_{\odot}$
yr$^{-1}$). They note that ``it is very unlikely that an accretion
disk mainly contributes to the pre-outburst luminosity''. It is
evident from our photographic photometry that there were no strong
emission lines in the SED of the progenitor which might disturb
the plain continuum in such a way as those in the SED of the
remnant. But we could not establish any contribution of Balmer
absorption or discontinuity of an A0 star due to absence of
$U$-band observations of the progenitor.

We determined the mean $<B>$ magnitude for V445 Pup to be 14<sup>m</sup>.30.
Taking into account the A0V energy distribution and the Galactic
reddening of $E(B-V) = 0.51$ (Iijima \& Nakanishi 2008), we find
$(B{-}V)_0 = 0.00$ which corresponds to the star's reddened values
$B{-}V = 0^m.51$ and $<V> = 13^m.79$. With the interstellar reddening
of 0.51 and the distance of 5 kpc, we have $M_V = -1^m.28$. With the
bolometric correction of --0.25 determined for an A0-type star by
Straizys (1982), we have $\log L/L_{\odot} = 2.54$. This value is
essentially lower than that of 4.34 derived by Woudt et al.
(2009). Even with their distance of 8.2 kpc, we find $M_V = -2^m.36$
and $\log L/L_{\odot} = 2.96$. These luminosity estimates are
still too bright to fit models of an accretion disk with a high
mass transfer rate. We think that the luminosity of the progenitor
was overestimated by Woudt et al. (2009), mostly due to inferring
additional circumstellar absorption by an equatorial disk/torus.
V445~Pup is an old-population object from its large radial
velocity, and its location in the CM diagram is about
$1\fmm5{-}2\fmm6$ above the RR~Lyrae horizontal-branch gap of the
globular cluster population.

On the other hand, the energy distributions of the V445 Pup
remnant are composite. In the $H$ and $K$ bands, this is mostly
continuum from  cool dust. Woudt et al. (2009) noted two emissions
of HeI in the 0.9--2.5 $\mu$m spectrum taken in January 2004; one
of them, at 1.0820 $\mu$m, is very strong. In their spectrum, the
continuum is present only beginning with 1.5~$\mu$m.

Woudt et al. (2009) wrote that V445 Pup was highly reddened in the
$JHK$ bands more than 8 years after the outburst. However, Fig.~2
in their paper shows that all the radiation in the $K$ band does
not belong to an A0 type star highly reddened by the ejected dust
envelope. It belongs to the dust envelope itself which consists of
two jet-like elongated lobes. No reddened stellar object (such as
an A0 type star) is located between the lobes. The $H$ and $K$
bands contained mostly the thermal component of $T = 250$~K, and
the intensity of this thermal component gradually declined. The
$J$ band contained the HeI $\lambda$2.0581 $\mu$m recombination
line with an equivalent width of approximately $-180$~\AA. The
behaviour of the flux in the $J$ band was different, it did not
decline. The figure shows some brightening of the remnant in the
$J$ band simultaneously with the decline in other infrared bands,
possibly due to appearance and strengthening of the HeI emission.

In the optical and near-infrared bands, the continuum of the
thermal dust component is not found, and we see the recombination
lines. In the 4500--7600 \AA\ spectrum taken with the VLT (Woudt
et al. 2009), four strong emission lines: [OIII] 4959 \AA, HeI
5875~\AA, HeI~7065 \AA, and the blend [OII] 7319/7330 \AA\
containing four components are present. The HeI and [OIII] lines
are spatially resolved. Therefore, our $BVR_CI_C$ magnitudes and
colours do not reflect temperature information, they are formed
only by line radiation accidentally hitting the photometric bands.
The [OIII] line is at opposite edges of the $B$ and $V$ passbands,
and the other lines are in the $R_C$ band. Our $BVR_CI_C$
observations show brightening of all emission lines in the time
ranges from JD 2453004 (r1 in Fig.~7) to JD 2453386 (r2) and
subsequent weakening from JD 2453386 (r2) to JD 2454777 (r3).

The changes in the SED due to the outburst in 2000 are radical. We
do not see such a blue and bright star as before the outburst.
Most investigators, from Henden et al. (2001) to Woudt et al.
(2009), write that the binary is obscured by the carbon dust shell
ejected in the outburst. However, we do not see such a bright and
blue but reddened star either in the deep IR spectra or in the
energy distributions. Does any radiation come from under the
shell? But there is the ionized gas, ions of [OII] and [OIII].
What is then the source of ionization and excitation?

The light curve of V445 Pup in the optical $V$ band presented in
Fig.~8 looks like a typical light curve of a dust nova having a
prolonged dip due to the dust obscuration. Good examples of
contemporary dust novae are V1419 Aql, V705 Cas, and V475 Sct. The
light curves of these stars can be found in the AAVSO database and
in our database at http://vgoray.front.ru/liststar.htm. For V445
Pup, such a dip continued approximately since JD $\sim 2452200$
till JD $\sim 2453500$. One may assume that, as in other dust
novae, the end of this dip and the subsequent star's rebrightening
are associated with the dispersion of dust and improvement of
visibility of the binary system's remnant. The dispersion of dust
leads to temporal strengthening of emission lines in the expanding
gaseous envelope. But after that, the tendency of weakening for
light from the gaseous envelope continued due to its expansion.
This process is typical of classical dust novae in the late stages
of outbursts.

\PZfig{8.5cm}{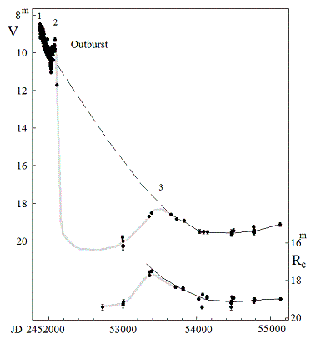}{The light curve of V445 Pup in
the $V$ band including the 2000 outburst. 1: The primary light
maximum; 2: rebrightening before the deep fading; 3: rebrightening
in the low state. The gray curve is a dip in the light curve due
to carbon dust obscuration. The dashed curve is a typical decay
without dust obscuration. The bottom light curve is the
post-outburst one in the $R_C$ band.}

\section{DISCUSSION}

There is a consensus that SNe Ia result from the explosion of a
carbon-oxygen (CO) white dwarf in a binary system (Meng \& Yang,
2010). The cited paper contains a good review of evolutionary ways
leading to the SN Ia explosion, with corresponding references to
the literature. The first theoretical studies of helium flashes on
the surface of a CO white dwarf were performed in the context of
SN Ia scenarios. Such dwarfs were members of common envelope
binary systems containing a low-mass red giant with a helium core
and mass transfer (Hachisu et al. 1989). Depending on the mass
transfer rate, helium burning on the surface of a white dwarf may
be stable or unstable, with flashes. As a result of helium burning
on the surface of a CO white dwarf, its mass gradually increases,
exceeds Chandrasekhar limit for the white dwarf mass, and
therefore the CO dwarf explodes as a thermonuclear runaway.
Hachisu et al. (1989) assume that the progenitor may be observed
as a symbiotic star (if the donor is a subgiant) or as an A--F
giant (if the donor is a red giant). In the latter case, the
mass-receiving component (mass gainer) has the photospheric radius
comparable with that of the donor.

Kato et al. (1989) discuss models of a helium nova outburst in a
compact binary with direct accretion of helium from a helium star
companion which fills its inner critical Roche lobe. In those
models, helium shell burning ignites when the density and the
temperature in the helium envelope reach critical values. The
envelope expands and forms dense stellar wind when it exceeds the
critical Roche lobe. When the helium burning ends, the star
returns to the same point in the luminosity--temperature ($L-T$)
diagram which was occupied by the progenitor. Helium flashes on
the CO white dwarf surface were not considered as a direct cause
of the SN explosion.

The appearance of a real helium nova, V445 Pup, showed that these
ideas were simplified: dust formation and asymmetry of ejecta were
not predicted. Dust forming and dissipation were not considered in
the paper by Kato et al. (2008) written after the V445 Pup
outburst, they considered only free-free-emission-dominated light
curves based solely on the optically thick wind theory.

Recent dynamic calculations by Guillochon et al. (2010) reveal the
mechanism of helium ignition on the surface of a white dwarf. With
the high accretion rates on the white dwarf surface that range
from $10^{-5}$ to $10^{-3}$ $M_\odot$ s$^{-1}$, the hot helium
torus forms, and the accretion stream fails to impact the white
dwarf surface. The mass of the torus accumulated during such an
event varies between 0.05 and 0.14 $M_\odot$. Due to velocity
difference between stream and torus, large-scale Kelvin--Helmholtz
instabilities arise along the interface between the two regions
which ``eventually grow dense knots of material that periodically
strike the surface of the primary, adiabatically compressing the
underlying helium torus''. The temperature of compressed material
increases above a critical temperature ($2\cdot10^9$ K), and then
triple-$\alpha$ reactions lead to detonation of the primary's
helium envelope. Moreover, Guillochon et al. (2010) show that
shock waves of the detonated envelope tend to concentrate in the
focal points within the CO core of the primary. The CO core of a
white dwarf can also detonate itself. This kind of detonation of
the CO core does not lead to a SN Ia, but such transient events
resemble dim type I SNe.

The cited authors used smoothed particle hydrodynamics (SPH)
simulations in their calculations. In such an event, the stability
of the system depends on the structure of primary and secondary
components. However, the particle approximation cannot fully
resolve the stream and the helium shell because they have small
masses as compared to the masses of the two components of the
binary. To increase the resolution, the authors used a hybrid
approach that combined both SPH- and grid-based methods.
Grid-based models predict that the donor will survive in such an
explosion.

\PZfig{8.5cm}{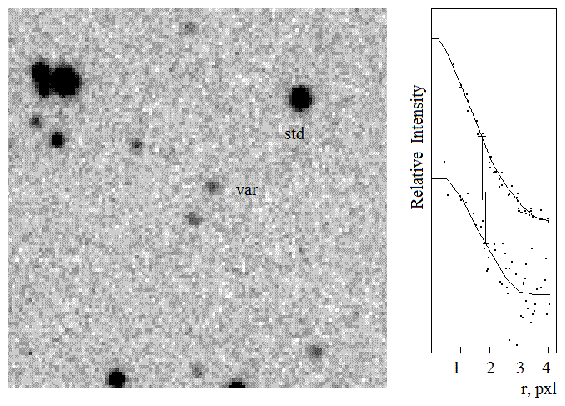}{An image of V445 Pup taken on 20
October 2009 with the SAO 1-m Zeiss telescope. This is a sum of
$V$ and $R_C$ frames. This image was obtained on the night with the
best seeing ever observed by us, FWHM = 1\farcs5. No elongation is
seen in the V445 Pup image (left). To the right, we compare pixel
brightness vs. pixel distance from the star center relations for
the V445 Pup remnant (bottom) and a nearby comparison star (top).
The curves are averaged star image profiles; the vertical lines
indicate star image radii at half intensity. }

Principally, Guillochon et al. (2010) presented models of merging
of a helium star and a CO white dwarf, i.e. a dynamically unstable
system in the last orbits before merging. In such an event, the
orbital period of the system may be unstable due to angular
momentum losses. Theoretical and observational studies of the
orbital period stability for a long time interval before the
merging event are needed to compare the V445 Pup explosion with
the merging model.

Fink et al. (2010) use dynamic calculations to investigate whether
an assumed surface helium detonation is capable of triggering a
subsequent detonation of the CO white dwarf core. Their
calculations were performed for a range of masses of white dwarf
cores and masses of helium shells on the white dwarfs' surfaces.
The authors find that secondary core detonations are triggered for
all of the simulated models, ranging in core mass from 0.810 up to
$1.385\, M_\odot$ with corresponding shell masses from 0.126 down
to $0.0035\, M_\odot$. The result of the calculations was the
following: ``as soon as a detonation triggers in a helium shell
covering a CO type white dwarf, a subsequent core detonation is
virtually inevitable''. For the case of the V445 Pup explosion
which began with a helium flash on the surface of the massive CO
white dwarf, the white dwarf was destroyed partly or totally. The
infrared bipolar nebulosity discovered using adaptive optics may
be the debris of the destroyed white dwarf. But V445 Pup is not a
dim SN Ia. Having an absolute magnitude $M_V$ between --5.9 and
--7.0 in the light maximum, it did not reach the absolute
magnitude level ranging between --13 and --14 for ultra-faint
supernovae (Smith et al. 2009), or so called ``supernova
impostors''.

What happened in the 2000 explosion event of V445 Pup? Our and
other studies established that the progenitor was a hot binary
system without a dense dust circumstellar disk. We do not see
absorption or emission of such a disk in the energy distribution
of the progenitor. The assumption of a cataclysmic binary with the
accretion disk is not probable because of high progenitor's
luminosity and an orbital light curve unusual for cataclysmic
variables. Indeed, a highly inclined cataclysmic variable has a
hump in its light curves caused by visibility of a hot spot on the
edge of the accretion disk; the hump is visible only during a half
of the orbital period. We assumed that the progenitor was a common
envelope binary with a luminous helium donor. The periodic
variability suggests a single star with a hot spot on its surface.
We know sdB + M type binaries with bright spots on the surfaces of
their cool components due to irradiation of ultraviolet flux from
an sdB star. However, it is not our case because we do not see a
cool companion, the companion was hot. The hypothesis of
irradiation of X-rays in a binary with a compact companion is
rejected due to absence of an X-ray source associated with the
progenitor.

Many observers detected dust formation during the outburst of V445
Pup and the strong dust absorption after the outburst. The
observations with adaptive optics reveal a highly collimated
bipolar dust and gas ejection located just in the plane of the
sky. The small inclination angle of ejection may confirm the
presence of an orthogonal dust structure closely aligned to the
line of sight and causing the strong extinction observed after the
outburst. In the case of merging components, such an orthogonal
structure might be a dust disk formed due to angular momentum
conservation, but the merging hypothesis needs orbital period
instability of the progenitor. Otherwise, this structure should be
radially expanding in the equatorial plane. Our optical photometry
reflecting changes of fluxes in HeI, [OII], and [OIII] emission
lines shows that the dust minimum finished in 2004, leading to
strengthening of these lines. Later on, emission line fluxes
continued to decrease. This means that the ionizing and exciting
radiation comes from outside the place of the explosion, from the
dust surroundings. However, no stellar object was seen in the
latest adaptive optics image taken in March, 2007.

The absence of a stellar source and the presence of an elongated
structure formed only from the ejecta in the frames taken with
adaptive optics suggest that components of the system lost their
mass and might be destroyed totally or partly. The interpretation
of the event as a helium flash is of no doubt. The dynamic theory
of such an explosion predicts detonation of the CO white dwarf
core and destruction of the white dwarf. Such an explosion will
undoubtedly result in the loss of the binary's common envelope.
The remnant may be the core of the helium donor which remains the
source of ionizing radiation. In the hypothesis of double
detonations, ejecta of the two detonations, the helium shell, and
the white dwarf core may be spatially separated.

We analyzed our CCD frames taken with the SAO 1-m reflector. The
best-resolution frame with the seeing FWHM = 1\farcs5 was obtained
on 20 October 2009. A part of this frame is shown in Fig.~9
(left). The star was faint, and to increase the signal in pixels,
we co-added our $V$ and $R_C$ frames. In the right panel, the
profile of a standard star (top) is compared to the profile of
V445 Pup (bottom). These are normalized dependences of pixel
counts on the distance between the pixel center and the center of
the star image localized by the method of star-profile dispersion
minimization. This distance is measured in the units of pixel
size, 0\farcs44. Both images are round, and the image of V445 Pup
is marginally resolved. No image elongation is seen, whereas the
expected elongation based on Fig.~6 by Woudt et al. (2009) should
be 2:1. One should note that the radiation in emission lines was
resolved by Woudt et al. (2009) with VLT spectra, and, very
probably, it coincides spatially with the lobes seen in the
infrared $K$ band.

In both cases, either CO type white dwarf detonation and
destruction or big mass loss from the helium donor, the future SN
Ia scenario for V445 Pup becomes impossible. The helium flash
accompanied with core detonation seems to be a single event in the
life of a white dwarf in the binary system with a helium star. It
does not look like a SN Ia. At least, it is too early to discuss
the SN Ia scenario for V445 Pup before the possible detection of
restored mass exchange in the binary system.

\section{CONCLUSIONS}

We found that V445 Pup progenitor was a binary system. The most
probable orbital period was $0\fday650654 \pm0\fday000011$. The
total variability amplitude was 0\fmm7 in the $B$ band, and the
total amplitude of the periodic variability component was 0\fmm4.
Solutions of the light curves with different periods are also
possible.

The shape of the progenitor's light curve suggests that it was a
common-envelope binary with a spot on the surface of the envelope
and with variable surface brightness.

The energy distribution of the progenitor can be successfully
fitted with that of a single A0V star with a small infrared
excess.

The shape of the light curve after the outburst suggests that the
dust absorption minimum finished in 2004. We do not see any
stellar remnant either in the optical range or in the infrared. No
stellar object, like a reddened A-type star, is visible. We argue
that V445 Pup was an event of double detonation, both of the
helium envelope of the CO white dwarf and of the CO white dwarf's
core, accompanied with the destruction of the common envelope of
the binary. In such an event, one of the components of V445 Pup
lost a part of its mass, and the other one might be totally
destroyed. Thus, V445 Pup, as a remnant of the outburst, cannot be
a progenitor of a future SN Ia.\\

\textit{Acknowledgments:} The authors thank A. Retter and A. Jones
for their observations of V445 Pup in outburst provided for our
analysis. We also used the Digitized Sky Survey, the SuperCOSMOS
Sky Surveys, the United States Naval Observatory (USNO) Image and
Catalog Archive, the USNO B1.0 astrometric catalog, and the All
Sky Automated Survey (ASAS) data. The information from deep sky
surveys is very useful for understanding the nature of
astrophysical objects and phenomena. E.A.B., A.V.Zh., and V.P.G.
thank Russian Foundation for Basic Research (RFBR) for financial
support through grant and No.~07-02-00630.
S.Yu.Sh. is grateful to the RFBR for grants No.~08-02-01229 and
No.~09-02-90458, to the Slovak Academy of Sciences for the VEGA
grant 2/0038/10, and to the Russian Ministry of Education and
Science for the grant RNP-2906.

\references

Ashok, N.M., \& Banerjee, D.P.K., 2003, {\it Astron. \&
Astrophys.}, {\bf 409}, 1007

Deeming, T.J., 1975, {\it Astrophys. \& Space Sci.}, {\bf 36}, 137

Fink, M., Roepke, F.K., Hillebrandt, W., et al., 2010, {\it
Astron. \& Astrophys.}, {\bf 514}, A53.

Glasner, A., \& Livne, E., 2002, AIP Conf. Proc., {\bf 637}, 124

Guillochon, J., Dan, M., Ramirez-Ruiz, E., \& Rosswog, S., 2010,
{\it Astrophys. J.}, {\bf 709}, L64

Hachisu, I., Kato, M., \& Saio, H., 1989, {\it Astrophys. J.},
{\bf 342}, L19

Henden, A.A., Wagner, R.M., \& Starrfield, S.G., 2001, {\it IAU
Circ.}, No.~7730

Iben, I., Jr., \& Tutukov, A.V., 1994, {\it Astrophys. J.}, {\bf
431}, 264

Iijima, T., \& Nakanishi, H., 2008, {\it Astron. \& Astrophys.},
{\bf 482}, 865

Kato, M., Saio, H., \& Hachisu, I., 1989, {\it Astrophys. J.},
{\bf 340}, 509

Kato, M., Hachisu, I., Kiyota, S., \& Saio, H., 2008, {\it
Astrophys. J.}, {\bf 684}, 1366

Kato, T., \& Kanatsu, K., 2000, {\it IAU Circ.}, No.~7552

Lafler, J. \& Kinman, T.D., 1965, {\it Astrophys. J., Suppl.
Ser.}, {\bf 11}, 216

Liller, W., 2001, {\it IAU Circ.}, No.~7561

Lynch, D.K., Russell, R.W., \& Sitko, M.L., 2001, {\it Astron.
J.}, {\bf 122}, 3313

Lynch, D.K., Rudy, R.J., Mazuk, S., et al., 2004, {\it IAU Circ.},
No.~8278

Meng, X. \& Yang, W., 2010, {\it Astrophys. J.}, {\bf 710}, 1310

Neckel, Th., Klare, G., \& Sarcander, M., 1980, {\it Astron. \&
Astrophys., Suppl. Ser.}, {\bf 42}, 251

Platais, I., Kozhurina-Platais, V., Zacharias, M.~I., \&
Zacharias, N., 2001, {\it IAU Circ.}, No.~7556

Smith, N., Ganeshalingam, M., Chornock, R., et al, 2009, {\it
Astrophys. J.}, 697, L49

Straizys, V., 1982, {\it Zvezdy s defitsitom metallov (Metal
Deficient Stars)}, Vilnius: Mokslas (in Russian)

Terebizh, V.Yu., 1992, {\it Analiz vremennykh ryadov v astrofizike
(Analysis of Time Series in Astrophysics)}, Moscow: Nauka (in
Russian)

Wagner, R.M., Foltz, C.B., \& Starrfield, S.G., 2001, {\it IAU
Circ.}, No.~7556

Williams, R.E., 1992, {\it Astron. J.}, {\bf 104}, 725

Wooden, W.H., II, 1970, {\it Astron. J.}, 75, 324

Woudt, P.A., Steeghs, D., Karowska, M., et al., 2009, {\it
Astrophys. J.}, 706, 738

\endreferences

\begin{table}%1
  \caption{Colours and magnitudes of comparison stars measured by A. Henden for VSNET}
  \smallskip
  \begin{center}
\begin{tabular}{rcccccccc}
\hline
\\
Star \#&$V$&$B-V$&$V-R$&$R-I$&$\Delta V$&$\Delta (B-V)$&$\Delta (V-R)$&$\Delta (R-I)$\\
\\
\hline
\\
1 &13.351&0.619&0.367&0.355&  8& 12& 11& 13\\
2 &13.851&1.565&0.917&0.828& 11& 31& 14& 11\\
3 &16.378&0.627&0.347&0.546&101&156&158&189\\
4 &15.707&1.479&0.704&0.831& 55&149& 71& 60\\
5 &14.799&0.586&0.362&0.375& 22& 36& 32& 37\\
6 &12.917&0.468&0.322&0.346&  6&  8&  8&  9\\
7 &13.887&0.533&0.334&0.322& 11& 16& 16& 18\\
8 &14.608&0.532&0.329&0.326& 21& 31& 32& 39\\
9 &14.727&0.721&0.446&0.393& 25& 41& 35& 40\\
10&14.558&0.418&0.235&0.208& 20& 28& 32& 45\\
11&14.394&2.324&1.317&1.337& 15& 97& 17& 11\\
12&13.932&1.431&0.817&0.776& 11& 31& 14& 12\\
13&15.190&1.208&0.737&0.731& 36& 76& 46& 39\\
14&14.676&1.210&0.719&0.750& 23& 50& 28& 21\\
15&14.032&1.148&0.664&0.688& 12& 27& 16& 14\\
16&14.062&1.447&0.812&0.802& 12& 37& 16& 13\\
17&14.066&0.355&0.201&0.251& 12& 17& 19& 24\\
18&15.557&1.001&0.604&0.504& 45& 91& 62& 64\\
19&16.082&0.765&0.265&0.495& 74&124&118&140\\
20&15.577&0.723&0.409&0.408& 47& 77& 69& 80\\
21&16.563&0.534&0.407&0.425&113&169&169&198\\
22&17.151&0.781&0.548&0.922&211&354&313&309\\
\\
\hline
\end{tabular}
\end{center}
\end{table}

\begin{table}%2
 \begin{center}
  \caption{Moscow $B$ magnitudes of V445 Pup}
  \smallskip
  \begin{tabular}{lllllllll}
  \hline
  \\
JD hel. & $B$ &$\sigma$ $B$& Plate& &JD hel. & $B$ &$\sigma$ $B$  & Plate    \\
-2400000&     &         & \ \ No. & &-2400000&     &         & \ \ No.   \\
\\
\hline
\\
40541.555 & 14.13 & 0.10 & A6833  & &44257.381 & 13.95 & 0.09 & A13697    \\
41274.543 & 14.32 & 0.02 & A7949  & &44261.369 & 14.04 & 0.08 & A13722    \\
41274.594 & 14.38 & 0.10 & A7950  & &44290.357 & 14.34 & 0.11 & A13749    \\
41356.357 & 14.13 & 0.12 & A7962  & &44315.238 & 14.18 & 0.07 & A13779    \\
41356.390 & 14.15 & 0.15 & A7963  & &44584.501 & 14.16 & 0.09 & A14157    \\
42719.569 & 14.59 & 0.23 & A10830 & &44672.252 & 14.35 & 0.09 & A14192    \\
42867.242 & 14.61 & 0.11 & A11009 & &44673.271 & 14.34 & 0.06 & A14202    \\
42870.243 & 13.97 & 0.15 & A11047 & &44905.601 & 14.36 & 0.06 & A14665    \\
42871.241 & 14.64 & 0.06 & A11059 & &44987.408 & 14.00 & 0.06 & A14691    \\
42872.242 & 13.98 & 0.08 & A11071 & &44988.411 & 14.39 & 0.08 & A14702    \\
42873.250 & 14.35 & 0.09 & A11083 & &44989.379 & 14.05 & 0.08 & A14717    \\
43160.394 & 14.09 & 0.02 & A11582 & &45054.243 & 14.36 & 0.07 & A14841    \\
43161.413 & 14.40 & 0.08 & A11593 & &45326.527 & 14.16 & 0.10 & A15269    \\
43163.444 & 14.31 & 0.05 & A11602 & &45327.464 & 14.11 & 0.07 & A15275    \\
43167.415 & 14.21 & 0.08 & A11615 & &45376.350 & 14.07 & 0.10 & A15327    \\
43190.348 & 14.32 & 0.08 & A11861 & &45396.265 & 14.40 & 0.07 & A15346    \\
43435.589 & 14.07 & 0.06 & A12370 & &45407.241 & 14.19 & 0.10 & A15356    \\
43850.498 & 14.28 & 0.07 & A12957 & &45642.578 & 14.13 & 0.04 & A15993    \\
43851.501 & 14.33 & 0.06 & A12963 & &45699.455 & 14.41 & 0.04 & A16076    \\
43865.406 & 14.55 & 0.07 & A12975 & &45703.433 & 14.39 & 0.02 & A16089    \\
43898.371 & 14.53 & 0.07 & A13008 & &45733.340 & 14.43 & 0.10 & A16120    \\
43905.327 & 14.22 & 0.06 & A13026 & &45734.342 & 14.11 & 0.08 & A16127    \\
43933.298 & 14.19 & 0.05 & A13066 & &47208.316 & 14.47 & 0.06 & A18258    \\
43935.280 & 13.97 & 0.04 & A13088 & &47620.246 & 14.54 & 0.08 & A18836    \\
43962.241 & 14.37 & 0.09 & A13110 & &47835.576 & 14.32 & 0.06 & A19355    \\
43964.245 & 14.61 & 0.03 & A13123 & &          &       &      &           \\
\\
\hline
\end{tabular}
\end{center}
\end{table}

\begin{table}%3
  \caption{Sonneberg $B$ magnitudes of V445 Pup}
  \smallskip
  \begin{center}
  \begin{tabular}{@{}lllllllll@{}}
  \hline
  \\
JD hel. & $B$ &$\sigma$ $B$  & Plate & &JD hel. & $B$ &$\sigma$ $B$  & Plate  \\
-2400000&     &          & \ \ No.   & &-2400000&     &          & \ \ No.\\
\\
\hline
\\
45779.321 & 14.23 & 0.24 & GC5336    & &46826.403 & 14.54 & 0.13 & GC7618/19/20\\
46006.624 & 14.30 & 0.11 & GC5821    & &46827.400 & 14.05 & 0.13 & GC7657      \\
46036.581 & 14.72 & 0.09 & GC5872    & &46827.416 & 14.33 & 0.08 & GC7658/59   \\
46059.505 & 14.46 & 0.12 & GC5910/11 & &46828.357 & 14.27 & 0.13 & GC7695/96   \\
46093.444 & 14.14 & 0.13 & GC5962/63 & &46828.420 & 14.47 & 0.11 & GC7698/99   \\
46113.374 & 14.66 & 0.09 & GC6043/47 & &46828.454 & 14.53 & 0.16 & GC7701/02   \\
46116.392 & 14.12 & 0.12 & GC6083/86 & &46851.344 & 14.22 & 0.19 & GC7771/72/73\\
46385.614 & 14.07 & 0.10 & GC6603    & &47566.376 & 14.28 & 0.23 & GC8869      \\
46387.611 & 14.02 & 0.15 & GC6645/46 & &47860.599 & 14.33 & 0.19 & GC9448/49   \\
46440.464 & 14.11 & 0.19 & GC6691/92 & &47861.609 & 14.46 & 0.10 & GC9477/78   \\
46733.639 & 14.24 & 0.08 & GC7425    & &47864.571 & 14.01 & 0.07 & GC9527/28   \\
46737.649 & 14.18 & 0.08 & GC7442/43 & &47943.357 & 14.54 & 0.12 & GC9541/42   \\
46763.582 & 14.41 & 0.05 & GC7526/27 & &48271.476 & 14.29 & 0.10 & GC9930/31   \\
46763.651 & 14.48 & 0.14 & GC7531    & &48273.485 & 14.34 & 0.08 & GC9973/74   \\
46768.607 & 14.52 & 0.05 & GC7568    & &48274.467 & 14.19 & 0.12 & GC10000/01  \\
46768.617 & 14.70 & 0.14 & GC7569    & &          &       &      &             \\
\\
\hline
\end{tabular}
\end{center}
\end{table}

\begin{table}%4
  \caption{Archive photographic magnitudes of V445 Pup
  from Digital Sky Survies}
  \smallskip
  \begin{center}
  \begin{tabular}{@{}lllllll@{}}
  \hline
  \\
JD hel. & $Mag$   &$\sigma$ & Band & DSS &  Emulsion & Plate \\
-2400000&         &         &      &     &  + Filter & \ \ \ No.\\
\\
\hline
\\
34397.736 & 14.06 &  0.08 & $B$ & POSS & 103aO         & SO843 \\
34397.815 & 13.18 &  0.08 & $R$ & POSS & 103aE         & SE842 \\
44333.908 & 14.29 &  0.12 & $B$ & UKST & IIIaJ + GG395 & 5815\\
45723.085 & 13.08 &  0.05 & $I$ & UKST & IV~N + RG715  & 8998\\
46772.806 & 13.40 &  0.12 & $R$ & ESO  & IIIaF + RG630 & 6704 \\
49750.077 & 13.37 &  0.08 & $R$ & UKST & IIIaF + OG590 & 16496\\
\\
\hline
\end{tabular}
\end{center}
\end{table}

\begin{table}%5
 \begin{center}
  \caption{CCD observations of V445 Pup after the outburst}
  \smallskip
  \begin{tabular}{@{}cccccc@{}}
  \hline
  \\
JD hel. & $B$ & $V$ & $R$ & $I$ &  Source$^1$\\
-2400000& \\
\\
\hline
\\
52729.2123 &  20.76 &   --  &    19.45 &  --   & SO \\
52997.4136 &    --  &   --  &    19.34 & 19.99 & SO \\
52997.4219 &    --  & 19.82 &    19.25 & 19.87 & SO \\
53004.3862 &    --  & 20.01 &    19.15 &  --   & SO \\
53004.3908 &  20.45 & 20.23 &     --   &  --   & SO \\
53004.3949 &  20.50 & 20.29 &     --   &  --   & SO \\
53357.5675 &    --  &   --  &    17.56 &  --   & VA \\
53357.5697 &    --  &   --  &    17.76 &  --   & VA \\
53357.5741 &    --  &   --  &    17.75 &  --   & VA \\
53357.5989 &  19.34 & 18.69 &    17.76 &  --   & VA \\
53386.4143 &  19.86 & 18.50 &    17.53 &  --   & SO \\
53387.3999 &  19.58 & 18.51 &    17.52 & 17.71 & SO \\
53652.5755 &    --  & 18.58 &      --  &  --   & SO \\
53728.5565 &  19.58 & 18.84 &    18.34 &  --   & VA \\
53827.2144 &    --  & 18.90 &    18.48 &  --   & SO \\
53827.2172 &    --  & 18.93 &    18.41 &  --   & SO \\
54035.5993 &    --  & 19.49 &    18.99 & 20.07 & SO \\
54079.5279 &    --  &   --  &    19.43 &  --   & VA \\
54091.4894 &    --  & 19.52 &    18.75 &  --   & VA \\
54146.3261 &    --  &   --  &    18.86 &  --   & SO \\
54146.3112 &    --  & 19.53 &    18.92 &  --   & SO \\
54468.4335 &  20.41 & 19.49 &    19.20 &  --   & SO \\
54468.4365 &  20.31 & 19.67 &    19.42 &  --   & SO \\
54468.4395 &  20.31 & 19.57 &      --  &  --   & SO \\
54468.4425 &  20.23 & 19.63 &      --  &  --   & SO \\
54478.4129 &  20.62 & 19.52 &    18.89 &  --   & SO \\
54478.4175 &  20.59 & 19.52 &      --  &  --   & SO \\
54479.4058 &  20.63 & 19.62 &    19.09 &  --   & SO \\
54499.3535 &  20.44 & 19.43 &    18.94 &  --   & SO \\
54766.5960 &    --  & 19.54 &      --  &  --   & SO \\
54766.6014 &    --  & 19.46 &      --  &  --   & SO \\
54767.5998 &    --  & 19.26 &    19.07 &  --   & SO \\
54774.5488 &    --  & 19.57 &    19.12 &  --   & SO \\
54776.5397 &  20.29 & 19.53 &    19.01 &  --   & SO \\
54778.5388 &  20.17 & 19.42 &    19.00 &  --   & SO \\
55125.6070 &    --  & 19.12 &    18.98 &  --   & SO \\
\\
\hline
\end{tabular}
\end{center}

\small{$^1$SO: Special Astrophysical Observatory, 1-m Zeiss
telescope; VA: SAI Crimean Station, 0.6-m Zeiss telescope.}
\end{table}

\newpage

\begin{table}%6
 \begin{center}
  \caption{Periods and frequencies revealed}
  \smallskip
  \begin{tabular}{@{}llcllll@{}}
  \hline
  \\
$P$& $\sigma~P$ & Parameter &$f$ (c/d) & Remark \\
\\
\hline
         &\\
Deeming  &          & $Ampl./2$&       & \\
         &          &          &       & \\
2.134469 & 0.00011  & 0.217 & 0.468501 & $f_0$\\
1.871862 & 0.00009  & 0.217 & 0.534227 & $1-f_0$\\
0.679704 & 0.000012 & 0.205 & 1.471229 & $1+f_0$\\
0.650637 & 0.000011 & 0.204 & 1.536956 & $2-f_0$\\
0.404207 & 0.000004 & 0.185 & 2.473977 & $2+f_0$\\
0.393764 & 0.000004 & 0.182 & 2.539658 & $3-f_0$\\
\\
\hline
         &\\
L--K     &          & $\theta$          \\
         &          &       &         & \\
0.650654 & 0.000011 & 0.718 & 1.536914& $2-f_0$\\
1.359423 & 0.000048 & 0.734 & 0.735606& d.w.$^2$\\
1.301269 & 0.000044 & 0.744 & 0.768481& d.w.\\
0.808410 & 0.000017 & 0.787 & 1.236996& d.w.\\
0.679686 & 0.000012 & 0.852 & 1.471269& $1+f_0$\\
0.404207 & 0.000004 & 0.844 & 2.471214& $2+f_0$\\
\\
\hline
\end{tabular}
\end{center}
\end{table}

{\small $^2$d.w.: a double-wave light curve.}

\end{document}